\documentstyle{mn}

\newif\ifAMStwofonts
\ifoldfss
  \ifCUPmtlplainloaded \else
    \NewTextAlphabet{textbfit} {cmbxti10} {}
    \NewTextAlphabet{textbfss} {cmssbx10} {}
    \NewMathAlphabet{mathbfit} {cmbxti10} {} % for math mode
    \NewMathAlphabet{mathbfss} {cmssbx10} {} %  "   "    "
  \fi
  \ifAMStwofonts
    \ifCUPmtlplainloaded \else
      \NewSymbolFont{upmath} {eurm10}
      \NewSymbolFont{AMSa} {msam10}
      \NewMathSymbol{\upi}     {0}{upmath}{19}
      \NewMathSymbol{\umu}     {0}{upmath}{16}
      \NewMathSymbol{\upartial}{0}{upmath}{40}
      \NewMathSymbol{\leqslant}{3}{AMSa}{36}
      \NewMathSymbol{\geqslant}{3}{AMSa}{3E}

    \fi
  \fi
\fi % End of OFSS

\ifnfssone
  \newmathalphabet{\mathit}
  \addtoversion{normal}{\mathit}{cmr}{m}{it}
  \addtoversion{bold}{\mathit}{cmr}{bx}{it}
  \newmathalphabet{\mathbfit} % math mode version of \textbfit{..}
  \addtoversion{normal}{\mathbfit}{cmr}{bx}{it}
  \addtoversion{bold}{\mathbfit}{cmr}{bx}{it}
  \newmathalphabet{\mathbfss} % math mode version of \textbfss{..}
  \addtoversion{normal}{\mathbfss}{cmss}{bx}{n}
  \addtoversion{bold}{\mathbfss}{cmss}{bx}{n}
  \ifAMStwofonts
    \ifCUPmtlplainloaded \else
      \UseAMStwoboldmath
      \makeatletter
      \new@mathgroup\upmath@group
      \define@mathgroup\mv@normal\upmath@group{eur}{m}{n}
      \define@mathgroup\mv@bold\upmath@group{eur}{b}{n}
      \edef\UPM{\hexnumber\upmath@group}
      \new@mathgroup\amsa@group
      \define@mathgroup\mv@normal\amsa@group{msa}{m}{n}
      \define@mathgroup\mv@bold\amsa@group{msa}{m}{n}
      \edef\AMSa{\hexnumber\amsa@group}
      \makeatother
     \mathchardef\upi="0\UPM19
      \mathchardef\umu="0\UPM16
      \mathchardef\upartial="0\UPM40
      \mathchardef\leqslant="3\AMSa36
      \mathchardef\geqslant="3\AMSa3E
    \fi
  \fi
\fi % End of NFSS release 1

\ifnfsstwo
  \DeclareMathAlphabet{\mathbfit}{OT1}{cmr}{bx}{it}
  \SetMathAlphabet\mathbfit{bold}{OT1}{cmr}{bx}{it}
  \DeclareMathAlphabet{\mathbfss}{OT1}{cmss}{bx}{n}
  \SetMathAlphabet\mathbfss{bold}{OT1}{cmss}{bx}{n}
  \ifAMStwofonts
    \ifCUPmtlplainloaded \else
      \DeclareSymbolFont{UPM}{U}{eur}{m}{n}
      \SetSymbolFont{UPM}{bold}{U}{eur}{b}{n}
      \DeclareSymbolFont{AMSa}{U}{msa}{m}{n}
      \DeclareMathSymbol{\upi}{0}{UPM}{"19}
      \DeclareMathSymbol{\umu}{0}{UPM}{"16}
      \DeclareMathSymbol{\upartial}{0}{UPM}{"40}
      \DeclareMathSymbol{\leqslant}{3}{AMSa}{"36}
      \DeclareMathSymbol{\geqslant}{3}{AMSa}{"3E}
    \fi
  \fi
\fi % End of NFSS release 2

\ifCUPmtlplainloaded \else
  \ifAMStwofonts \else % If no AMS fonts
    \def\upi{\pi}
    \def\umu{\mu}
    \def\upartial{\partial}
  \fi
\fi

\input{epsfig.sty}

\title{Hysteresis in the Light Curves of Soft X-ray Transients}
\author[T. J. Maccarone \& P. S. Coppi]
       {Thomas J. Maccarone \\
        Scuola Internationale Superiore di Studi Avanzati, via Beirut, n. 2-4, Trieste, Italy, 34014 
	\newauthor
	Paolo S. Coppi\\
	Department of Astronomy, Yale University, P.O. Box 208101, New Haven CT USA 06520-8101}
	
\date{}

\pagerange{\pageref{firstpage}--\pageref{lastpage}}
\pubyear{}

\begin{document}

\maketitle

\label{firstpage}

\begin{abstract}

Using PCA data from the Rossi X-Ray Timing Explorer (RXTE), we track
the spectral states of the neutron star transient system Aql X-1
through a complete outburst cycle. We find a hard-to-soft state
transition during the very early, rising phase of the outburst and
show that there is a hysteresis effect such that the transition back
to the hard state occurs at a luminosity $\sim $ 5 times {\it lower}
than the hard-to-soft transition. This hysteresis effect rules out the
propeller mechanism as the sole cause of state transitions in Aql
X-1. Assuming the propeller mechanism only operates at a luminosity
equal to or below that of the observed soft-to-hard transition
requires that the magnetic field of Aql X-1 be less than $7\times
10^7$ Gauss, the lowest neutron star field known to date.  To compare
the state transition behavior of Aql X-1 with that found in transient
black hole systems, we use RXTE ASM data to compute hardness-intensity
diagrams for four black hole candidate transients where the ASM data
should also give us state information throughout much of the outburst
cycles. In all four systems, we find evidence for a hard-to-soft state
transition during the rising outburst phase and for the source staying
in a soft state down to much lower luminosities during the declining
phase, i.e., a hysteresis effect.  This similarity suggests a common
origin for state transitions in low magnetic field neutron star and
black hole systems, and the hysteresis effect rules out the ``Strong
ADAF Principle'' for determining the state of an accretion disc. We
discuss the general implications of these observations for current
models of state transitions.  We note the contrast to previous
observations of the non-transient systems Cygnus X-1 and X-3 which do
not show a hysteresis effect.
\end{abstract}

\begin{keywords}
accretion,accretion discs -- X-rays:binaries -- X-rays:individual:Aql X-1
\end{keywords}

\section{Introduction}
Soft X-ray transients (SXTs) represent a class of X-ray binaries which
undergo occasional, perhaps quasi-periodic outburst cycles during
which their luminosities rise by factors of $\sim100-10000$ (in the
case of neutron star systems) to factors of $\sim10^6$ or more (in the
case of black hole systems).  Their spectral states also change during
these luminosity transitions.  At the lowest luminosities for which
the spectrum can be measured well, their continua generally look to be
a hard ($\frac{dN}{dE}\sim E^{-1.5}$) power law, with a cutoff
typically between 50 and 500 keV - the low/hard state.  The low/hard
state emission mechanism is generally thought to be thermal
Comptonisation by hot electrons (e.g. Sunyaev \& Titarchuk 1980).  At
luminosities greater than $\sim$ a few percent of the Eddington
luminosity, the spectrum resembles, to first order, the standard
multi-temperature blackbody model of Shakura \& Sunyaev (1973) - the
high/soft state.  In some cases, a ``very high state'' exists where
the multi-temperature blackbody from the thin disc is seen at the same
time as a hard power law tail (albeit with a spectral photon index
typically around 2.5) that extends to $\gamma$-ray energies with no
apparent cutoff.  The radiation mechanisms in the very high state are
not as well understood as in the other states.  There also exists an
``off state'' at very low luminosities.  Current observational
evidence seems to favor the idea that the off state is merely an
extension of the low/hard state to very low luminosities (see
e.g. Corbel et al. 2000 - C00), but the data is relatively sparse.  For a
review of the properties of the different spectral states, see Nowak
(1995).

Past work has shown that the spectral state correlates with the system
luminosity (see e.g. the reviews by Tanaka \& Lewin 1995; Nowak
1995). In black hole systems, the spectral state transitions are
currently thought to be driven by the shift in accretion flow geometry
from a thin disc that extends to $\sim$ the innermost stable circular
orbit (high/soft state) to either a thin disc with a hole in it or a
thin disc with a hot corona above it (the low/hard state).  In this
low/hard state, the hard X-rays are then thought to be produced by
Compton scattering of these disc photons in a hot, quasi-spherical
flow inside this hole (e.g. Shapiro, Lightman \& Eardley 1976;
Ichimaru 1977; Rees et al. 1982; Narayan \& Yi 1994), a magnetically
powered corona above the disc (e.g. Nayakshin \& Svensson 2001), or
the hot base of a relativistic jet (e.g. C00; see also Markoff, Falcke
\& Fender 2001 who claim the X-rays are synchrotron radiation from
this jet).

To the extent that the accretion mode is determined only by the
presence of a deep gravitational potential well, these models may
apply to neutron star SXTs as well.  However, neutron stars may
possess significant magnetic fields and the ``propeller model'' may be
relevant in explaining the state transitions in some or all of these
systems (e.g. Lamb, Pethick \& Pines 1973; Zhang, Yu \& Zhang 1998;
Campana et al. 1998).  In the propeller model, the inner region is
kept empty of mass by the magnetosphere of the neutron star when the
magnetospheric radius is larger than the corotation radius, and hence
the magnetic pressure of the magnetosphere is larger than the gas' ram
pressure.  The neutron star then becomes very much like an isolated
radio pulsar and the X-ray luminosity is then provided by some
combination of a small fraction of the gas accreting along the
magnetic poles and the collision of the pulsar wind with the accreting
gas (Campana et al. 1998).

Until recently, the standard lore has held that the SXT outbursts
usually proceed directly from quiescence into the high/soft state,
perhaps going through a very high state first.  This assumption has
been recently contradicted by the observations of XTE J 1550-564
(Wilson \& Done 2001), XTE 1859+226 (Brocksopp et al. 2002) and of Aql
X-1 (Jain 2001), who have found hard states in the rising portions of
the outbursts.  These initial hard states are not very long lived, and
it is only due to the advent of the RXTE All-Sky Monitor, which has
greater sensitivity than previous all-sky monitors, to dedicated
optical monitoring programs (e.g. Jain 2001 and references within),
and to the rapid response of the pointed instruments on the Rossi
X-Ray Timing Explorer that detections of this brief hard state have
been possible.

In this paper, we discuss Rossi X-Ray Timing Explorer light curves
from the All Sky Monitor (ASM) and from the pointed Proportional
Counter Array (PCA) of several soft X-ray transients: the neutron star
system Aql X-1, and the black hole candidates 1748-288, 1859+226,
2012+381, XTE J1550-564, and GRO J1655-40 (see Liu, van Paradijs, \&
van den Heuvel 2001 and references within for discussion of source
identifications; see also Orosz et al. 2002 for a dynamical
confirmation that XTE J 1550-564 contains a black hole primary star;
Fillipenko \& Chornock 2001 for analogous work on 1859+226).  We show
that in all cases except 1655-40, the lightcurves seem to show hard
states in the rising and falling phases of the outburst and hysteresis
loops in diagrams of spectral hardness versus intensity.  We discuss
the implications for the different models of state transitions in
SXTs.  We note that the similarities between the outburst cycles could
imply a common origin for the state transitions of black hole and
neutron star SXTs and that a particular difficulty is presented for
propeller models as the sole mechanism for state transitions in
neutron star systems.

\section{Observations}
In general, hardness ratios of neutron star transients in the low/hard
state cannot be well constrained by the ASM on RXTE because the source
fluxes are too low.  Fortunately, a full outburst cycle has been
observed with the PCA for Aql X-1 by using ground-based optical
monitoring from the Yale 1m telescope (operated by the YALO consortium
- see Bailyn et al. 1999) to trigger the series of
target-of-opportunity X-ray observations we analyse here. (See Jain
2001 for a discussion of the optical observations).  Because the
full-scale outbursts in Aql X-1 occur roughly once per year, and
because the optical flux rises several days earlier than the X-ray
flux, such a monitoring program is feasible, effective, and necessary
in order to catch the rising phase of the X-ray outburst with pointed
instruments.  It is more difficult to get useful pointed observations
of typical black hole transient sources in all spectral states -
either they are new transients, in which case one does not know where
to point the monitoring optical telescope, or they are recurring ones,
but with substantially longer outburst cycles (typically $\sim$ 10
years).

Aql X-1 is also an especially useful source for testing propeller
models in neutron star systems because its spin frequency of 550 Hz
has been measured as the coherent quasi-periodic oscillation (QPO)
frequency in a Type I X-ray burst (W. Zhang et al. 1998).  Until
recently, there had been some controversy as to whether the coherent
QPO frequency represented the neutron star's spin frequency or twice
its spin frequency.  A series of 5 bursts from KS 1731-260 with
coherent QPOs was examined to look for oscillations at 1/2 and 3/2 of
the strongest coherent QPO frequency and an upper limit on the power
at these frequencies of 5\% of the power of the strongest frequency
was found (Muno et al. 2000), making it extremely unlikely that the
strongest frequency seen during the Type I bursts is actually a
harmonic (although see Abramowicz, Kluzniak \& Lasota 2001 for an
argument {\it against} the claim that the coherent kilohertz QPOs show
the rotational frequency of the neutron star).

For the black hole candidates, data from the ASM (see Levine et
al. 1996 for an instrument description) are used.  We take the ratio
of counts from 5-12 keV (band 3) to counts from 1-3 keV (band 1) and
construct a new hardness ratio.  Since the typical disc temperature
for a soft X-ray transient is about 1 keV (giving a spectral peak near
3 keV), the 1-3 keV band should be dominated by disc emission whenever
a strong disc is present and the 5-13 keV band should be well above
the cutoff energy for the disc component.  Thus this ratio should give
a good indication of whether the accretion flow is in the hard state
(where the $\nu F_\nu$ peak will be well above 10 keV) or in the soft
state (where the $\nu F_\nu$ peak will be below 5 keV).  An exception
must be made in the case of GRO J 1655-40.  In this source, the disc
temperature in the high/soft state can be quite large - up to a few
keV (e.g. Sobczak et al. 2000).  This can push the peak energy in the
disc spectrum up past 5 keV (i.e. into the hardest band on the ASM)
and can make any ratio of the count rates in two ASM bands a poor
indicator of the spectral state.  The ASM hardness ratios for GRO J
1655-40 seem to scale with luminosity (i.e. the source becomes harder
as it becomes brighter, because its disc temperature and disc flux are
correlated), regardless of spectral state, consistent with these
expectations, and thus indicating that the ASM cannot constrain the
ratio of disc to power law flux for this system.  For XTE J 1550-564,
which has undergone three outburst cycles since the beginning of the
RXTE mission (e.g. Jain et al. 2001; Smith et al. 2000) we plot the
results from only the earliest (late summer-early fall 1998) outburst,
which is qualitatively similar to the plot for the other two outburst
cycles.  We further restrict the data sets by including only those
days where the count rates in all three bands are positive (allowing
for the construction of hardness ratios) and the net count rate in all
three bands is greater than 1.5 counts per second.  The crude spectral
measures from the ASM and the rather large uncertainties in the
distances of the black hole systems will preclude accurate
measurements of the luminosities of these systems from being made, but
are sufficient for comparing relative fluxes for a given source.

The PCA data for Aql X-1 are extracted using the standard FTOOLS 5.0,
using the standard RXTE Guest Observer Facility recommendations for
screening criteria.  The observations used are from the May/June 1999
outburst (proposal IDs 40047 and 40049). They are then fit in XSPEC
11.0 using a thermal Comptonisation model (EQPAIR - see Coppi 1998 for
a description of the model).  The details of the spectral fitting
procedure and the results of the spectral fits will be described in a
follow-up paper (Maccarone \& Coppi, in preparation).  The essence of
the procedure that is relevant to this work is that (1) by fitting a
model, we may accurately estimate the flux from the source and (2) the
model returns as a fit parameter the ratio of luminosity in the
Comptonising electrons to the luminosity in the disc, which indicates
the spectral state of the system.

\section{Results}

For Aql X-1, we plot flux versus the ratio of Comptonising electron
luminosity in the corona to disc luminosity in Figure 1a.  Arrows
trace the outburst cycle from earliest time to latest time. To
facilitate comparisons with the black hole systems, we also plot, in
Figure 1b, the Aql X-1 data converted into ASM units.  We make this
conversion by dividing the ASM model counts by the Crab model counts
(as given in Wilms et al. 1999) in the two energy ranges (1-3 keV and
5-12 keV) then multiplying the resulting ratios by the Crab count
rates.  Since the Crab ratios do not strictly represent a response
matrix, slight systematic errors may be introduced in this way.
Additional systematic errors may be introduced by the extrapolation of
the best fitting model for Aql X-1 to energies below the range where
the fit was made.  Nonetheless, this figure should make it easier for
the reader to see the qualitative similarities between the black hole
and neutron star systems.  The figures look essentially the same with
one exception - the compactness monotonically decreases early in the
Aql X-1 outburst cycle, but the hardness increases for the first few
points.  This time period corresponds to an increase in the
best-fitting seed photon temperature which pushes the peak of the
accretion disc's spectrum from the lowest ASM channel to the middle
ASM channel.  When the temperature is low and the disc's spectrum is
peaking in the lowest channel, the ratio of counts in band three to
counts in band 1 is lower for a given ratio of disc luminosity to
corona luminosity than it does when the disc spectrum peaks in the
second ASM channel.

In Aql X-1, the observed transition from hard state to soft state
occurs at a flux level of $4.2-5.5 \times 10^{36}$ ergs/sec, assuming
a distance of 2.5 kpc (Chevalier et al. 1999), while the transition
from the soft state back to the hard state occurs at
$6.1-7.5\times10^{35}$ ergs/sec; that is to say, the ratio of
luminosity in the hard component to luminosity in the soft component
remains roughly constant from the onset of the soft state until the
return to the hard state a factor of 5-10 lower in luminosity.  The
possibility of hysteresis behavior in Aql X-1 has been pointed out
based on observations the system at a higher luminosity in the hard
state of one outburst cycle than in the soft state of another outburst
cycle (see e.g. Cui et al. 1998; Reig et al. 2000), but the effect is
shown here definitively for the first time.  Some additional previous
evidence for hysteresis has been seen in other sources - steadily
accreting neutron stars have shown signs of hysteresis in the
transitions between island and banana states (Muno, Remillard \&
Chakrabarty 2002), the candidate microquasar XTE 1550-564 shows brief
interludes where its spectral and variability properties seem to
resemble those of the very high state, but without its luminosity
rising above the typical luminosity in the high/soft state (Homan et
al. 2001), the black hole candidate GX 339-4 shows a very similar
hysteresis effect in its state transitions between the low/hard and
high/soft states (Grebenev et al. 1991; Miyamoto et al. 1995; Nowak
1995; Nowak, Wilms \& Dove 2002), and several persistently bright low
mass X-ray binaries containing black holes have shown different
luminosities for the hard-to-soft and soft-to-hard state transitions
(SHS).

The results from the All-Sky monitor work on the black hole systems
are plotted in Figure 2.  We note that analogously to the case of Aql
X-1, the hardness ratios of the black hole systems remain roughly
constant over a drop by a factor of $\sim$ 10 in count rate.  Like the
inferred ASM rates for Aql X-1, all four figures span 3 decades in
count rate and two decades in hardness ratio.  Since the count rates
for 1859+226 and 2012+381 are similar to the count rates for Aql X-1,
these three sources are all plotted for the same range of hardness and
count rate values, so their curves are not centered in the graphs.
The only major difference is that Aql X-1 is somewhat harder than the
two black hole sources at all count rates, but this offset is roughly
constant as a function of count rate and we cannot be sure whether it
is a real effect or is an artefact of the process by which we have
inferred the ASM count rates from the PCA.  Because the hardness ratio
is roughly constant, the conversion from count rate to luminosity
should also remain roughly constant.  Often the transition from
soft-to-hard cannot be seen because by the time the transition is
made, the flux level has dropped below the threshold of 1.5 counts per
second, and the hardness ratios are hence unreliable due to poor
statistics. Thus we cannot be certain that the state transition that
occurs is to the ``low/hard'' state rahter than the ``off'' state, but
as noted above and in C00, there is no evidence for substantial
differences in properties between the low/hard and off/state; the off
state has generally been defined based on source luminosities being
lower than instrumental thresholds and not based on any spectral
transition or rapid luminosity change from the low state.  Arrows
trace the outburst cycles from earliest time to latest time.  The
errors are left unplotted for XTE J 1748-288; because this source lies
in a high background region toward the Galactic Center (Revnitsev et
al. 2000), the errors for this source are quite large and plotting
them severely confuses the figure.  Thus conclusions about 1748-288
should be weighed accordingly.  We plot the results nonetheless
because they seem qualitatively similar to the other three sources for
which good ASM lightcurves can be made.

\begin{figure*}
\centerline{\epsfxsize=7 cm \epsfysize=9 cm \epsfbox{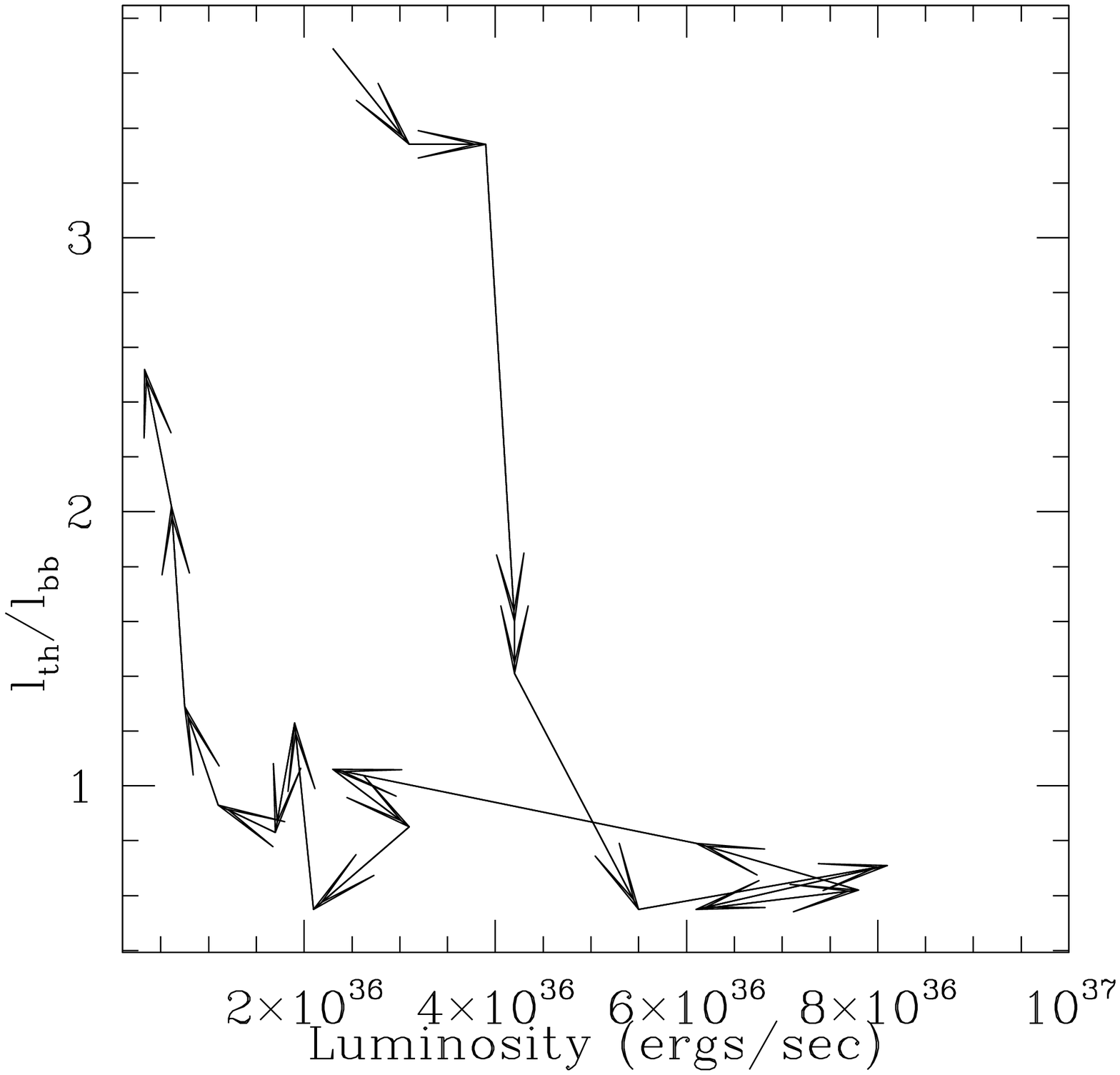}\epsfxsize=7 cm \epsfysize=9 cm  \epsfbox{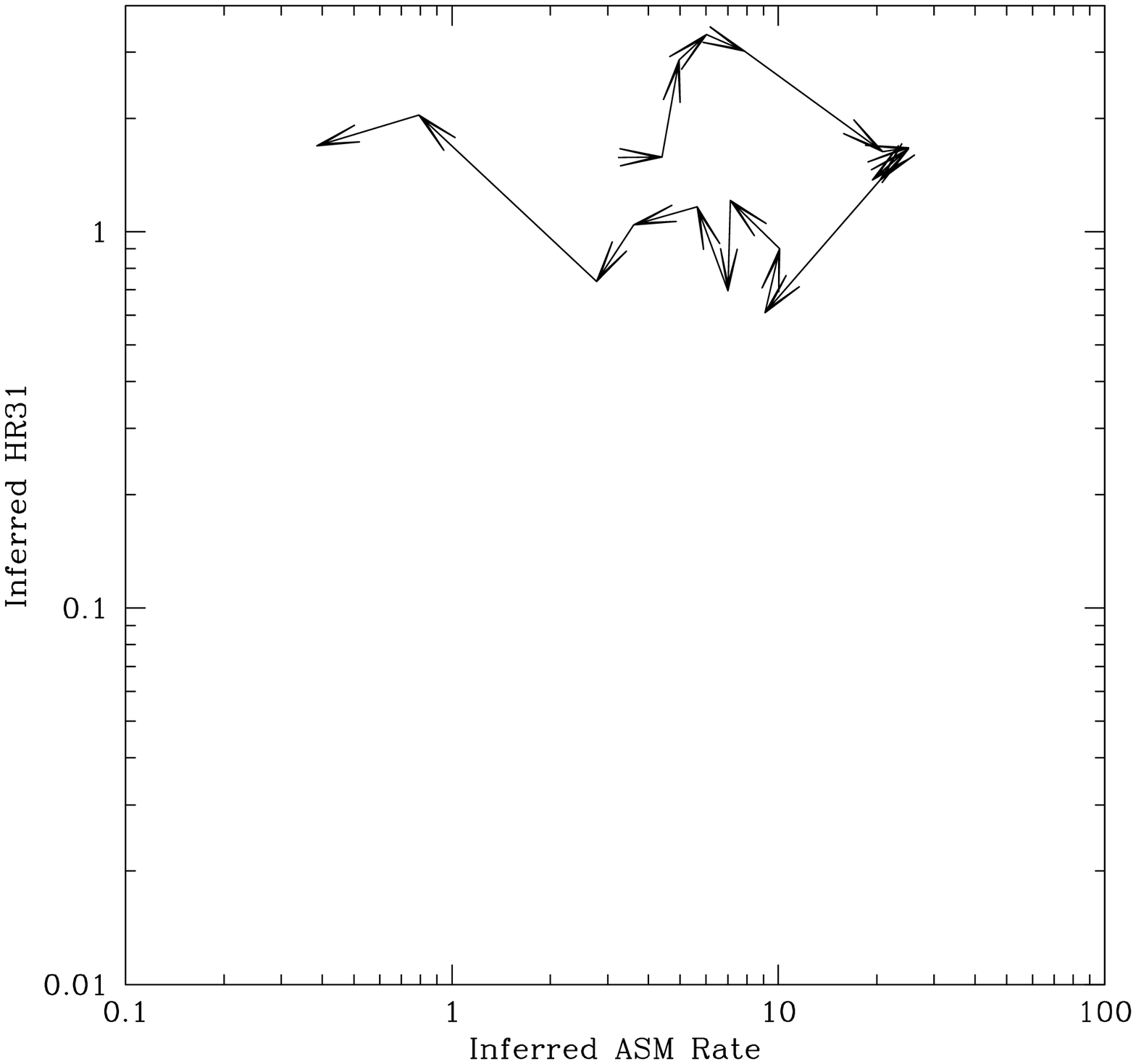}}
\caption{Left: The luminosity of Aql X-1 in ergs/second plotted versus
the ratio of the luminosity of the corona to the luminosity of the
disc.  The errors are left unplotted in order to avoid confusing the
plot, but are typically about 0.1 for $\ell_{th}/\ell_s$ and about 1\%
for the luminosity.  Right: The inferred ASM count rate (all bands)
plotted versus the inferred ASM hardness ratio (Counts from 5-12
keV)/(Counts from 1-3 keV) for Aql X-1.  The conversion is done to
allow an easier comparison with the black hole data in Figure 2.  In
both plots, the arrows trace the outburst as a function of time.}
\end{figure*}

\begin{figure*}
\centerline{\epsfxsize=7 cm \epsfbox{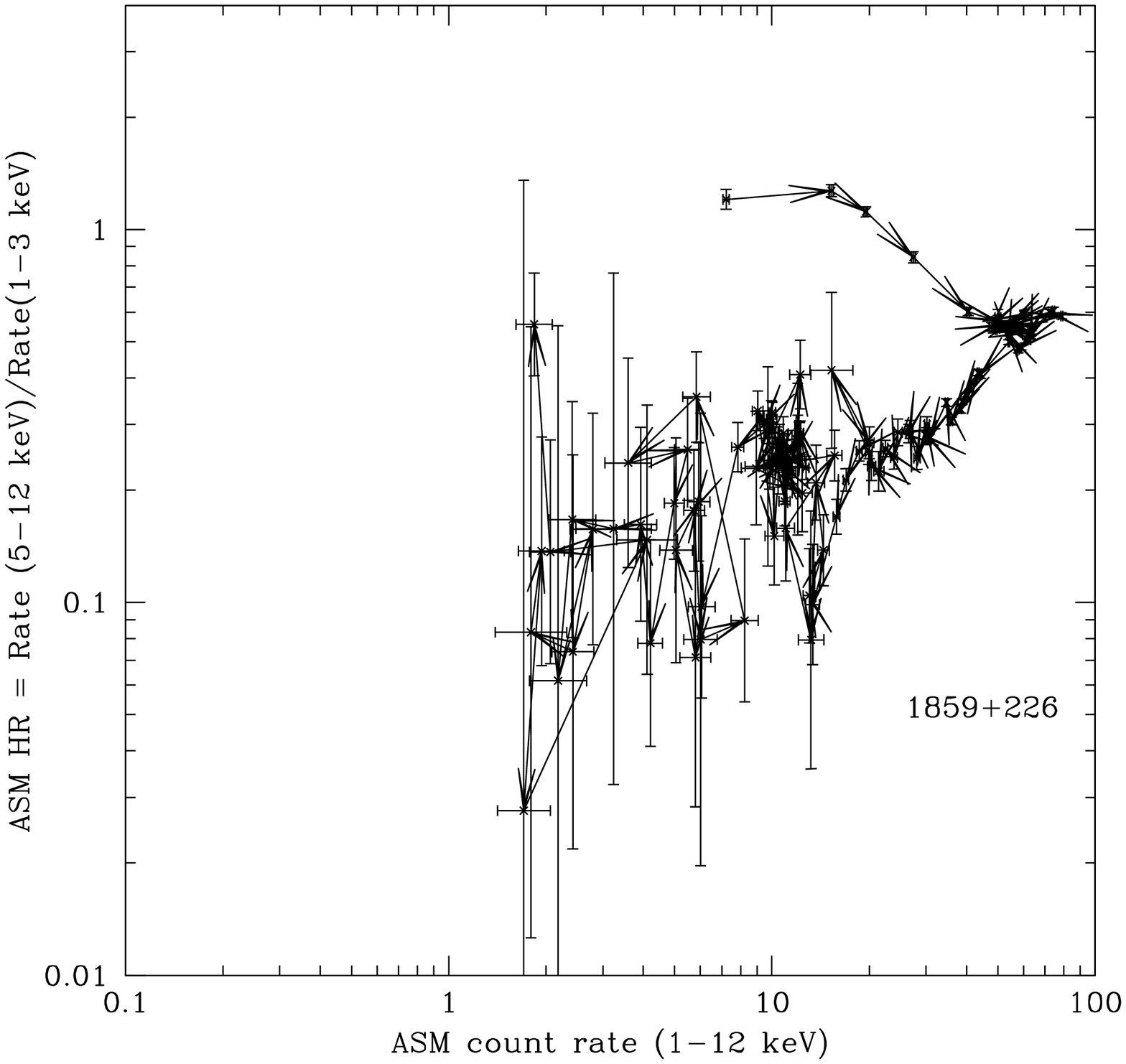}\epsfxsize=7 cm \epsfbox{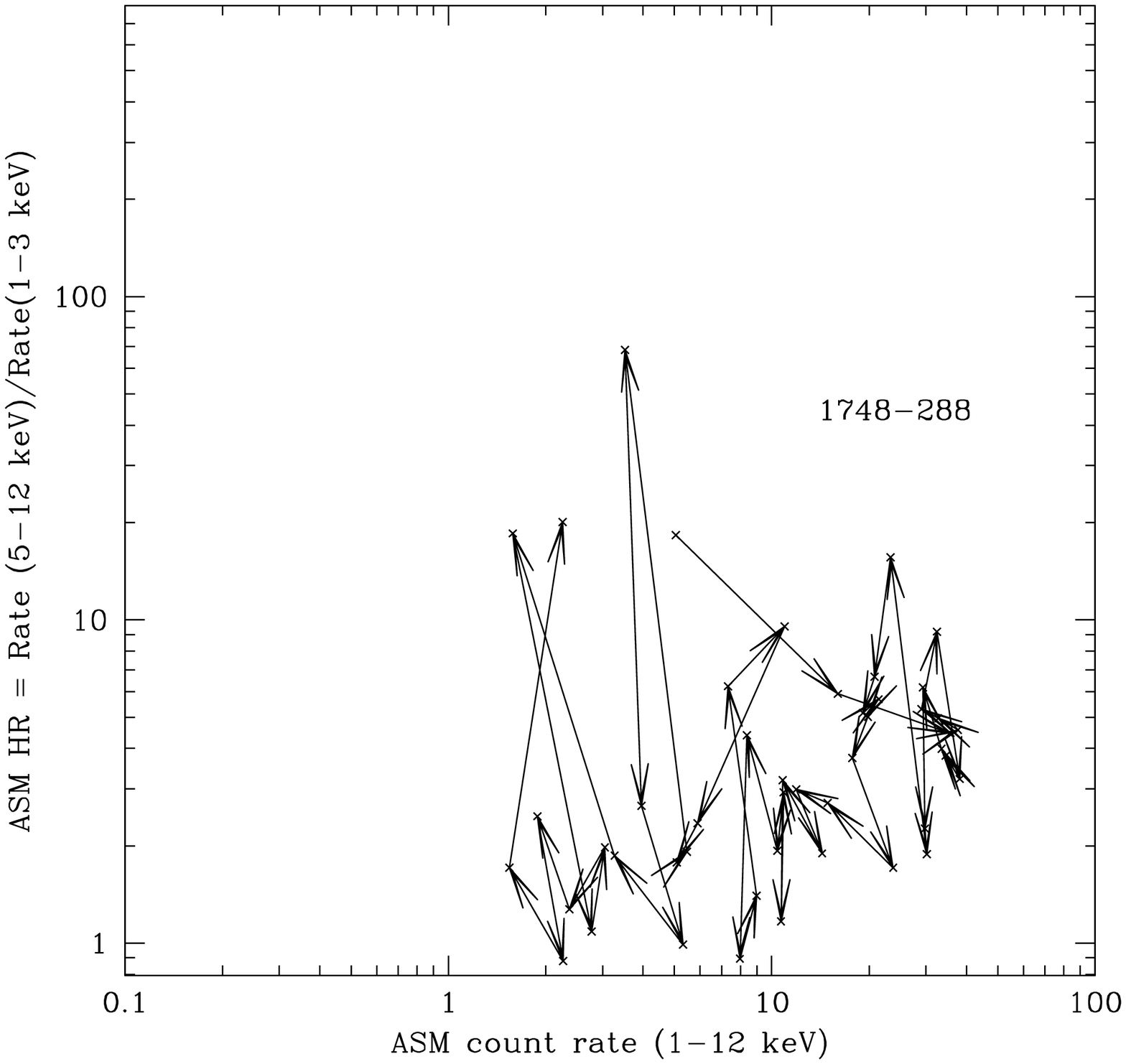}}
\centerline{\epsfxsize=7 cm \epsfbox{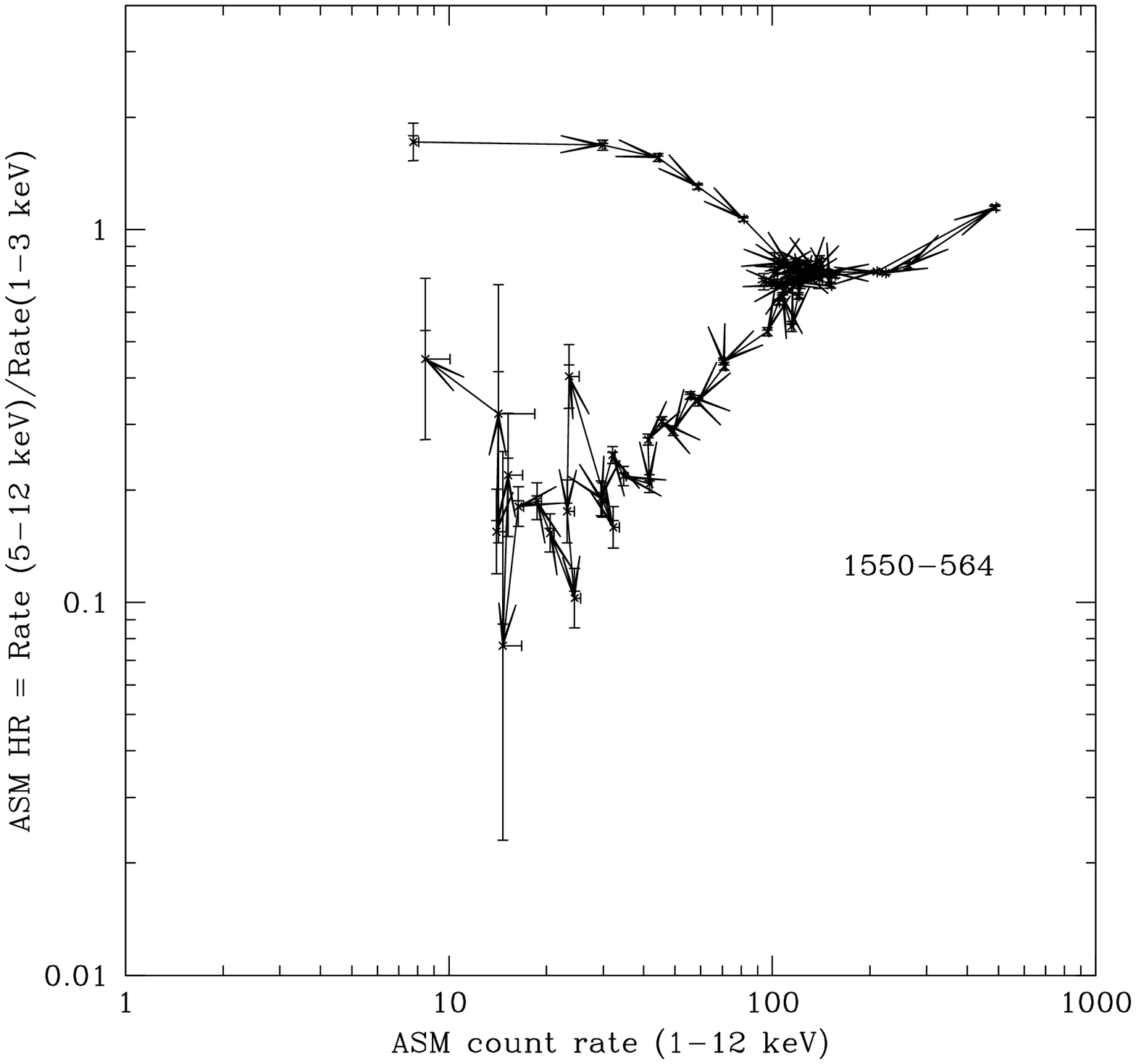}\epsfxsize=7 cm \epsfbox{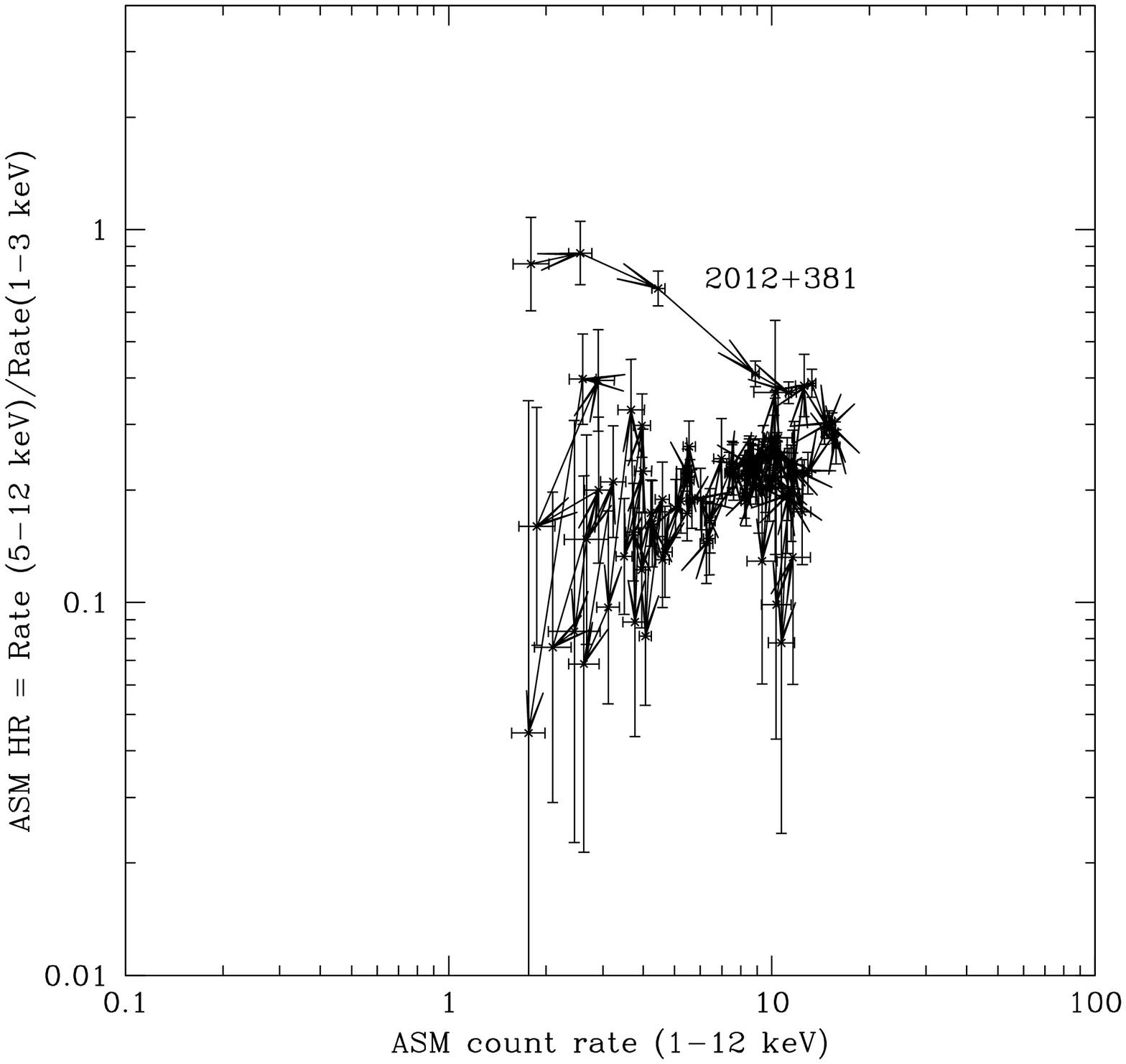}}
\caption{ASM count rate versus hardness ratio for the four candidate
black hole transients.  The arrows trace the outbursts as a function
of time.}
\end{figure*}

\section{Discussion}

\subsection{Connection with previous hysteresis observations}
Hysteresis results have previously been noted for several X-ray binary
black hole systems that are not classical transients because they
never enter the off state (Smith, Heindl \& Swank 2002 - SHS).  The
power law indices of these sources were compared with their photon
fluxes.  For the low mass X-ray binary systems, the hardening of the
power law components was found to lag behind the drop in brightness.
Such behaviour is qualitatively similar to what we have observed in
the classical transients.

The high mass X-ray binaries in the SHS sample, Cygnus X-1 and Cygnus
X-3 do not fit this trend.  It has been shown that for both these
systems, the RXTE count rate and spectral index track one another with
only short time lags (SHS).  SHS thus suggest that the hard and soft
components of the spectra are caused by two different accretion flows
(a picture similar to that invoked by van der Klis 2001 to explain
properties of quasi-periodic oscillations), and that the difference
between the two Cygnus systems and the other X-ray transients is
caused by the high mass nature of the mass donor.  Since wind driven
accretion tends to form smaller accretion discs, disturbances should
pass through the discs more quickly.  Changes in the hard X-ray
accretion flow (the corona) occur quickly in all systems, while
changes in the soft X-ray flow (the disc), occur on a timescale of
order the viscous timescale of the accretion disc.  The accretion
disks are larger in the low mass systems than in the high mass
systems, so the viscous timescales might be expected to be longer in
the low mass systems.

We note, however, that the ``soft'' state of Cygnus X-1 is not a
typical soft state, as it has a rather substantial hard X-ray tail,
probably related to Comptonisation off non-thermal electrons (see
e.g. Gierlinski et al. 1999).  Furthermore, it occurs at a luminosity
probably not much higher than the low/hard state luminosity (see
e.g. Frontera et al. 2001), and the bolometric luminosity changes in
Cygnus X-1 are, in fact, never very large.  As we discuss below, the
hysteresis effect could in part reflect the radical, non-equilibrium
nature of a transient outburst, particularly during the rising
phase. Cygnus X-1 never experiences such violent changes.

We note that the case of Cygnus X-3 is likely to be unique.  Given
that the most likely distance estimate of 11.5 kpc implies Cyg X-3 has
a minimum luminosity of $\sim 2 \times 10^{38}$ ergs/sec and that its
strong radio flares correspond with strong X-ray flares (Mioduszewski
et al. 2001), and that archival Green Bank Interferometer data show
that the radio flux from Cygnus X-3 in the 2.3 GHz and 8.3 GHz bands
never drops below 20 mJy for more than a few days at a time over the
$\sim$ 4 yrs of public data, we conclude that Cyg X-3 is likely to be
in the very high state at all times.  Its spectral state changes are
thus more likely to be like those in GRS 1915+105 (see e.g. Belloni et
al. 2000) than those of Cygnus X-1 or of the soft X-ray transients.
It would thus be beneficial for us to observe some high mass X-ray
binaries which go through the classical set of state transitions
before evaluating whether the ``two accretion flow'' scenario is
correct.  At present no such systems are known in the Galaxy, and for
theoretical reasons, it {\it may} be that no wind-driven soft X-ray
transients can exist.

\subsubsection{Comparison with GX 339-4 Radio/X-ray Results}
A phenomenological similarity exists between the results found here
and the hysteresis diagram of radio flux versus X-ray flux for GX
339-4 found in C00.  That plot shows that the radio flux is strong and
correlated with the X-ray flux during the hard state, falls as the
X-ray flux rises in the transition state, and is unobservable during
the high/soft state.  The X-ray flux then falls during the soft state
while the steady radio flux remains quenched.  The source sometimes
then enters an off-state where it is undetectable in X-rays and in
radio, but sometimes proceeds directly back into the low luminosity
end of the hard state.  The hard-to-soft transitions always occur at
higher luminosities than the soft-to-hard or soft-to-off transitions.
It has been noted elsewhere that the hard state luminosities of GX
339-4 sometimes exceed its soft state luminosities (e.g. Nowak, Wilms
\& Dove al. 2002; Kong et al. 2002).  While GX 339-4 shows all the
spectral states of X-ray transients, it is not usually considered a
classical soft X-ray transient because it frequently spends long
periods of time in the hard state and does not always go into the
off/quiescent state in between high state episodes.  Still the fact
that its outburst cycles also show hysteresis loops suggest that the
phenomenon is present in X-ray transients of all kinds.  Furthermore,
that the radio emission is completely quenched in the soft state
suggests that no persistent outflow is occurring in the soft state.
This provides additional, albeit indirect, evidence that the wind in
the soft state of Aql X-1 is likely to be weak if present since strong
outflows generally show some radio emission.  The analogies between GX
339-4 and the more standard transient sources also underscore the need
for relatively deep radio monitoring data on the soft X-ray
transients.  While many of these sources are monitored in the radio by
the VLA (and were monitored by the Green Bank Interferometer), the
observations so far have generally not been deep enough to make
detections (M.Rupen, private communication).  We note that we have
tentative evidence for a 0.2 mJy detection of Aql X-1 as it enters the
low/hard state in the 2002 outburst (Maccarone et al. 2002, in prep).
It is still not clear whether these observations indicate that the
jet/outflow is supplying the X-rays, as suggested by C00 (see also
Markoff, Falcke \& Fender 2001) or that the jet has a power
proportional to the X-ray luminosity and only turns on when the
accretion flow is geometrically thick (see e.g. Meier 2001).

\subsection{Implications for Propeller Models and the Magnetic Field 
of Aql X-1}
The similarities between the outburst cycles of the neutron star
system Aql X-1 and the black hole systems are quite striking.  The
simplest conclusion is that the same mechanism drives the state
transition in both accreting black holes and accreting neutron stars.
This in turn casts serious doubt on propeller models for the state
transitions in Aql X-1, which to date has been one of the strongest
candidates for showing this effect (Zhang, Yu, \& Zhang 1998 - ZYZ;
Campana et al. 1998).  According to ZYZ (1998), the propeller effect
becomes important at a critical luminosity, determined by the magnetic
field and the rotation rate of the neutron star, properties which
remain nearly constant throughout a single outburst cycle.  Under
these assumptions the propeller model can be unequivocally ruled out
as the sole cause of state transitions based on these observations.
In the derivation of the critical propeller luminosity (Lamb, Pethick,
\& Pines 1973 - LPP), it is shown that it is the ram pressure and
hence the accretion rate itself that is the critical factor, not the
luminosity.  Thus if one posits a strong wind inside the
magnetospheric radius only after the soft state has started, then one
can keep a high ram pressure at the magnetospheric radius while
reducing the flux of mass that reaches the surface of the neutron star
and hence reducing the luminosity of the accretion flow.  Still no
current models exist to explain how such an outflow would occur, and
even given such a model, the difficulty of explaining the very similar
black hole state transitions with a qualitatively different model
would remain.  In fact, most current theoretical work suggests that
outflows should be more important in the low/hard state than in the
high/soft state (see for example, Blandford \& Begelman 1999).

Having established that standard propeller effects can cause only the
soft-to-hard transitions (and then only possibly), we can place an
upper bound on the magnetic field strength by finding the critical
magnetic field needed for propeller effects to be important:
\begin{equation}
B_9=P_{-2}^{7/6}M_{1.4}^{1/3}R_6^{-5/2}L_{x,36}^{1/2},
\end{equation}
where $L_{x,36}$ is the X-ray luminosity in $10^{36}$ ergs/second,
$B_9$ is the surface magnetic field in units of $10^9$ Gauss, $P_{-2}$
is the spin period in units of 10 milliseconds, $M_{1.4}$ is the
neutron star mass in units of 1.4 solar masses, and $R_6$ is the
neutron star's radius in units of 10 km (LPP; ZYZ).  Setting $L_x$ to
$6.1\times 10^{35}$ ergs/sec (the lower bound for the soft-to-hard
state transition), the radius to 10 km, the mass to 1.4 $M_\odot$, and
the period to 1.8 milliseconds (as given by the 550 Hz quasiperiodic
oscillation), we find that the magnetic field must be less than
$7\times10^7$ Gauss.  In fact, the magnetic field would only be this
large if the soft-to-hard state transition is cause by the propeller
effect; if it is not caused by propellers, then the field could be
arbitrarily small.  Explaining the hard-to-soft state transition with
propeller effects would require a magnetic field of about
$2\times10^8$ Gauss.

An additional problem apart from the necessity of a wind in the soft
state arises from explaining both transitions with the propeller
effect - the magnetic field of the neutron star will then exceed its
equilibrium value, in contradiction to generally accepted models of
millisecond pulsar evolution (e.g. Srinivasan 1995).  If one replaces
the instantaneous luminosity in equation (1) with the mean luminosity,
then the equilibrium magnetic field (i.e. the magnetic field where the
spin-up due to accretion and the spin-down due to magnetic braking are
in equilibrium) results (Bhattacharya 1995).  Given the mean
luminosity for Aql X-1 of $\sim5\times10^{35}$ ergs/second as
estimated from the ASM count rates, a 1.8 millisecond period from the
550 kHz QPO, and the canonical values for the other parameters, the
equilibrium magnetic field would be $6\times10^7$ Gauss, far less than
the $2\times10^8$ Gauss field needed to cause the hard-to-soft
transition. 

A quick timing analysis has shown that there are no coherent
pulsations at 550 Hz in the RXTE data down at the level of $\sim$ 2\%
in the two low state observations that occurred after the soft-to-hard
transition.  This provides potentially strong evidence that the
propeller effects are not likely to cause the state transitions in Aql
X-1.  This observation is not definitive, though, because the
intrinsically narrow QPO from matter falling along the magnetic poles
may be broadened in time by the bulk Comptonisation that causes the
X-ray spectrum in the propeller state to be hard.  The magnetospheric
radius of Aql X-1 is at least 500 kilometers, since the state
transition luminosity is no more than $7.5\times10^{35}$ ergs/sec.
Since the bulk Comptonisation will occur mostly outside the
magnetospheric radius (and should not be in the extreme Klein-Nishina
limit where forward scattering dominates), the initially coherent
pulses will be smeared out by the Comptonisation.  Since the light
travel time though the magnetosphere is of over the pulse period, this
smearing effect could sufficient to make the coherent QPOs
unobservable in the hard X-ray bands of RXTE.  A future timing mission
with better soft X-ray coverage, such as the proposed timing mission
for XEUS (Barret et al. 2002) could make a more definitive statement
based on the lack of pulsed emission.

Propeller effects might still be important at lower accretion rates
such as those which put the source in quiescence (e.g. Menou \&
McClintock 2001) and hence with Aql X-1's having a lower magnetic
field than that claimed by ZYZ (1998) and by Campana et al. (1998).
In fact, given that the mean $L$ for Aql X-1 is very close to the $L$
at the transition from soft state to hard state, (and hence if the
source is in equilibrium, the propeller effect's critical mass
accretion rate is close to the soft-to-hard transition accretion rate)
the propeller effect may set in very shortly after the transition from
soft state to hard state, or it may even cause the soft-to-hard state
transition to occur at a slightly higher accretion rate than would
occur in the absence of a magnetic field.  If, however, the neutron
star is still spinning up, then the magnetic field must be lower than
the equilibrium value of $6 \times 10^7$ G.  Such a magnetic field
value would be lower than any other measured neutron star magnetic
field (see Cheng \& Zhang 2000 for a list of pulsar magnetic fields),
which should not be too surprising, since few pulsars have frequencies
as fast as the 550 Hz frequency of Aql X-1.  

Magnetic screening by accreted material on the surface of a neutron
star has been suggested as a mechanism for finding an inferred
magnetic field lower than the actual surface value, but has been found
to be important only for accretion rates above $\sim1\%$ of the
Eddington luminosity (Cumming, Zweibel \& Bildsten 2001).  Since the
luminosities here (both for the soft-to-hard state transition and for
the mean luminosity) are well below that value, magnetic screening is
unlikely to be important.  If the mean luminosity of Aql X-1 in the
last $\sim$ 100-1000 years has been substantially (i.e. about 5 times)
higher than the average over the RXTE mission lifetime, magnetic
screening could be important.

\subsection{Implications for Advection Dominated Flows}

Outburst cycles in the advection dominated accretion flow picture are
also often explained in terms of disc instability models (e.g. Menou
et al., 2000).  The triggering of an outburst due to the disc
instability and the change from an advection dominated flow to a thin
disc need not happen at the same accretion rate.  Despite the claim in
the standard picture that the state changes depend only on the
accretion rate (e.g. Esin, McClintock, \& Narayan 1997) it is likely
that in the ADAF picture, some sort of hysteresis would occur.
Advection (at low luminosities) occurs when the mass density is too
low for electrons and protons to exchange energy efficiently through
Coulomb interactions.  When the accretion flow is geometrically thick,
the mass density and hence the interaction rate between electrons and
protons are reduced compared to when the accretion flow is
geometrically thin.  Thus the critical accretion rate below which the
disc must drop to become a geometrically thick ADAF should be smaller
than the critical accretion rate above which the ADAF must collapse
into a thin disc.  In fact, the work of Zdziarski (1998) shows that
the luminosity of the state transitions should be approximately $0.15
y^{3/5}\alpha^{7/5} L_{EDD}$, where
$y=(4k_BT/m_ec^2)Max(\tau,\tau^2)$.  Thus barring a large increase in
$\alpha$ during the soft state, the hard-to-soft transition should be
expected to occur at a higher luminosity than the soft-to-hard
transition, since $y$ will be higher in the hard state than in the
soft state.

Three possible scenarios have been outlined for determining whether a
system will enter an advection dominated flow: the ``Strong ADAF
Principle'', which states that a system will enter an advection
dominated flow whenever such a solution is possible; the ``Weak ADAF
Principle'', which states that the ADAF will be chosen whenever it is
the only {\it stable} solution possible; and the ``Initial Condition
Principle'', which states that the initial conditions determine which
solution will be chosen (Narayan \& Yi 1995; Svensson 1999).  The
observations of hard spectra at the high accretion rates seen in the
rising portion of the low/hard state imply that advection dominated
flows are possible (if they do, indeed describe the hard states) at
the luminosities where the system is seen in the soft state in the
decaying phase of the outburst.  This provides a clear rejection of
the Strong ADAF principle.  It is likely that the system is stable in
the soft state where the luminosity is dropping, because the
luminosity is changing rather slowly here.  It {\it may} be unstable
in the rising portion where the luminosity is changing rapidly, and
may hence be pushed into a spectral state it would not enter if the
luminosity changes were slow.  Thus either the ``Weak ADAF Principle''
may apply, if ADAF solutions are stable when the disc luminosity is
changing rapidly, but thin disc solutions are not stable to rapid
luminosity changes or the Initial Condition Principle may apply.  In
fact, the standard ADAF may be unstable itself (Blandford \& Begelman
1999).  Whether the specific hot adiabatic accretion flow in the hard
state is the ADAF picture or one of the scenarios with outflows or
convection is beyond the scope of this paper.

\subsection{Disc evolution and evaporation?}
A possible picture to explain how the initial conditions affect the
spectral state is that when the accretion rate shuts off at the end of
the outburst, the system stays in a disc-dominated flow until the disc
is evaporated into the corona (e.g. Mineshige 1996; Cannizzo 2000;
Meyer, Liu \& Meyer-Hofmeister 2000; Dubus, Hameury \& Lasota 2001).
The observed hysteresis effect may then be a result of the time
between when the evaporation dominates over disk inflow (i.e. when the
system begins filling the corona with gas) and when the corona
dominates the total energetics of the system (i.e. when the corona
becomes filled).  The fraction of the accretion flow that is pumped
into the corona and the disc may depend only on the accretion rate,
but the state of the system will depend also on the recent history of
the accretion flow.  The inner disc will fill in as it accretes matter
from the disc at larger radii, but the inner disc will empty by
dumping matter into the corona.  Since the matter follows a different
geometric path on its way into and out of the disc, hysteresis might
be expected.  This picture has already had some success in explaining
the fast-rise exponential decay light curve profiles for the outbursts
of X-ray transients (Cannizzo 2000).  Numerical modeling is currently
underway to determine whether this picture can match the quantitative
details of the observations (J. Cannizzo, private communication).  In
essence, in this picture, this disc can fill in much more rapidly than
it can be evacuated by evaporation or other means.  Such a scenario
might also be able to explain the results seen in Cygnus X-1, where,
as noted by SHS, the disc is likely to be smaller, and where the
luminosity changes are also smaller, indicating that the system is not
as severely perturbed from the steady state.  In essence, the class of
models where the rate of disc filling and evacuation determines the
properties of the accretion flow solution is a realisation of the
``Initial Condition Principle'' - when the system is changing rapidly
as in the outburst, it may enter into non-steady state regions of
phase space where an ADAF is supported at relatively high accretion
rates.  When the system is changing slowly as in the decline, or in
systems like Cygnus X-1 where the luminosity is nearly constant over
the outburst, these regions of phase space are not accessed.

Other evidence for initial conditions affecting the accretion flow on
rather long timescales has been seen in the parallel tracks of
luminosity versus QPO frequencies seen in several low mass X-ray
binaries (van der Klis 2001).  In this picture, the only important
initial condition is the time averaged luminosity.  The key feature is
that there are two accretion flows, one radial and one disclike.  The
radial flow is assumed to be a constant fraction of the disc flow, but
to be averaged over the accretion rate in a large annulus of the disc.
The radial flow propagates inwards more quickly than does the disc
flow, so changes in the overall accretion rate are manifested in the
radial flow before they are manifested in the disclike flow, and
mathematically, it can be represented as a time average of the future
disclike accretion rate.  The total luminosity is then the disc
luminosity plus the radial flow's luminosity.  The quasi-periodic
oscillation frequencies then are assumed to vary as a function of the
instantaneous accretion rate divided by the mean luminosity.  Thus
parallel tracks are formed because the disc luminosity varies much
more quickly than the radial flow luminosity.  The increase of
frequency with luminosity on short timescales comes from the fact that
only the disc luminosity changes on short timescales.  The parallel
offsets of the tracks come from changes in the radial component's
contribution to the luminosity.

Our data challenge this specific two-flow picture.  In the Aql X-1
data, where the soft-to-hard state transition is well observed, the
state transition is rapid compared to the decay; after about 20 days
of steadily dropping luminosity in the high/soft state, the spectrum
goes from a high optical depth and a low compactness to a low optical
depth and high compactness in two days.  If the radial flow, which
makes up the hard component, represents a time-average of the future
disc luminosity, then the transition from the soft state back to the
hard state should not be rapid for an outburst which decays slowly in
luminosity and one would expect the source's spectrum to harden quite
gradually.  This picture has success in predicting the properties of
the quasi-periodic oscillations, however, and should not be rejected
too hastily. The best test of this model, tracking the QPO frequencies
through the outburst so that the time-averaged luminosity can be
measured directly and compared with the QPO frequencies is not easily
accomplished; the QPOs are not strong enough to be seen in all these
Aql X-1 observations, most likely because not all the proportional
counters are turned on for all the observations and some of the
exposures are rather short.  The scenario of van der Klis (2002) might
be modified by making the radial-to-disc accretion ratio a function of
the luminosity, which presumably becomes a steep function around the
state transition luminosity.  It then becomes essentially similar to
the disc evaporation picture described above, with the added feature
of explaining the quasi-periodic oscillation frequencies.

\section{Conclusions}

We find that for soft X-ray transients with a neutron star primary
(Aql X-1), and four soft X-ray transients with black hole candidate
primaries (certainly XTE J 1550-564, XTE J 1859+226, and XTE J
2012+381, and probably XTE J 1748-288), plots of spectral state versus
luminosity show hysteresis loops.  In all cases, hard states are seen
in the rising phases of the outburst cycle.  The luminosity of the
hard-to-soft transition is generally a factor of $\sim$ 5 or more
brighter than the soft-to-hard transition.  The observation for Aql
X-1 suggests that the turning off of the propeller effect cannot cause
the state transition from the hard state to the soft state.  The
magnetic field of the neutron star in Aql X-1 can be constrained to be
$<7\times10^7$ Gauss, a lower value than that for any other known
neutron star.  The similarities between the outburst cycles in neutron
stars and black hole candidates suggest a common state transition
mechanism.  Three promising possibilities are (1) the state transition
luminosity from an adiabatic accretion flow to a thin disk is higher
than the transition luminosity from a thin disk to an adiabatic flow
because interactions are more efficient in the thin disk where the
mean particle separation is smaller (2) during the rapid luminosity
rise, a geometrically thin accretion flow is not stable, so the
geometrically thick flow persists because the system is out of
equilibrium and (3) a time lag is present in the transition from thin
disk to geometrically thick accretion flow because the disk must be
evacuated or evaporated.  Given the present data and the present state
of theoretical work, we cannot distinguish among these possibilities.
The analogous hysteresis properties of the radio and hard X-ray fluxes
of GX 339-4 lends some credence to the idea that the effects of jets
may be of some importance.  Observations of the full series of
spectral states of additional neutron stars may to determine whether
the soft-to-hard state transitions may be routinely caused by
propeller effects, and hence whether the similarity between the mean
luminosity and the soft-to-hard transition luminosity is a coincidence
or is due to the soft-to-hard state transition's being caused by
propeller effects.  Better monitoring of soft X-ray transients at
radio wavelengths is also necessary to improve our understanding of
what role jets may play in the production of the hard X-rays.

\section{Acknowledgements}

We wish to thank Charles Bailyn, Mike Nowak, and Cole Miller for
useful suggestions.  We also wish to thank Raj Jain and Charles Bailyn
for sharing the results of their optical/X-ray monitoring campaign of
Aql X-1 in advance of publication.  We thank John Cannizzo for
pointing out the potential importance of disc evaporation.  We thank
the referee, Chris Done, for useful comments which have improved both
the quality and clarity of this paper, especially for pointing out the
possibility of non-equilibrium effects mattering in the rising hard
state.  This work is based on results provided by the ASM/RXTE teams
at MIT and at the RXTE SOF and GOF at NASA's GSFC.  This research has
made use of data obtained from the High Energy Astrophysics Science
Archive Research Center (HEASARC), provided by NASA's Goddard Space
Flight Center.

\label{lastpage}

\begin{thebibliography}{}
\bibitem{}Abramowicz, M.A., Kluzniak, W., \& Lasota, J.-P., 2001, A\&A, 374, L16
\bibitem{}Barret, D. et al., 2002, proceedings of the Workshop
"XEUS-studying the evolution of the hot universe" held in Garching,
March 11-13, 2002, eds. G. Hasinger, Th. Boller and A. Parmar
(astro-ph/0206028)

\bibitem{}Bailyn, C.D., Depoy, D., Agostinho, R., Mendez, R., Espinoza, J., \&
Gonzalez, D., 2000, BAAS, 195, 87.06

\bibitem{}Belloni, T., Klein-Wolt, M., M\'endez, M., van der Klis, M.  \& van
Paradijs, J., 2000, A\&A, 355, 271

\bibitem{}Bhattacharya, D., 1995, in {\it X-Ray Binaries}, eds. Lewin, van
Paradijs, \& van den Heuvel, Cambridge University Press : Cambridge

\bibitem{}Blandford, R.D. \& Begelman, M.C., 1999, MNRAS, 303, 1L

\bibitem{}Brocksopp, C., et al., 2002, MNRAS, 331, 765

\bibitem{}Campana, S. et al., 1998, ApJL, 499, 65

\bibitem{}Cannizzo, J.K., 2000, ApJL, 534, 35

\bibitem{}Cheng, K.S., \& Zhang, C.M., 2000, A\&A, 361, 1001

\bibitem{}Chevalier, C., Ilovaisky, S. A., Leisy, P., \& Patat, F., 1999, A\&A,
347, 51

\bibitem{} Coppi, P.S., 1998, ``The Physics of Hybrid
Thermal/Non-Thermal Plasmas,'' in {\it High Energy Processes in
Accreting Black Holes}, eds. J. Poutanen \& R. Svesson, ASP Conf
Ser. Vol. 161, p. 385 (astro-ph/9903158)

\bibitem{}Corbel, S., Fender, R.P., Tzioumis, A.K., Nowak, M., McIntyre, V.,
Durouchoux, P., \& Sood, R., 2000, A\&A, 359, 251 (C00)

\bibitem{}Cui, W., Barret, D., Zhang, S.N., Chen, W., Boirin, L., \& Swank, J.,
1998, ApJL, 502, 49

\bibitem{}Cumming, A., Zweibel, E. \& Bildsten, L., 2001, ApJ, 557, 958

\bibitem{}Dubus, G., Hameury, J.-M. \& Lasota, J.-P., 2001, A\&A, 373, 251

\bibitem{}Esin, A.A., McClintock, J.E., \& Narayan, R., 1997, ApJ, 489, 865

\bibitem{}Fillipenko, A.V. \& Chornock, R., 2001, IAUC No. 7644

\bibitem{}Frontera, F., et al., 2001, ApJ, 546, 1027

\bibitem{}Gierli\'nski, M., Zdziarski, A. A.,  Poutanen, J., Coppi, P.S., Ebisawa, K., \& Johnson, W. N., 1999, MNRAS, 309, 496

\bibitem{}Grebenev, S., Syunyaev, R.A., Pavlinskii, M.N. \& Dekhanov, L.A., 1991, Sov. Astron. Lett., 17, 413

\bibitem{}Homan, J., Wijnands, R., van der Klis, M., Belloni, T., van
Paradijs, J.,  Klein-Wolt, M., Fender,  R., M\'endez, M.,  2001, ApJS,
132, 377

\bibitem{}Ichimaru, S., 1977, ApJ, 214, 840

\bibitem{}Jain, R.K., 2001, Ph.D. Thesis, Yale University

\bibitem{}Kong, A.K.H., Charles, P.A., Kuulkers, E., \& Kitamoto, S., 2002,
MNRAS, 329, 588

\bibitem{}Lamb, F.K., Pethick, C.J., \& Pines, D., 1973, ApJ, 184, 271

\bibitem{}Levine, A.M.,  et al. 1996, ApJL, 469, L33

\bibitem{}Liu, Q.Z., van Paradijs, J., \& van den Heuvel, E.P.J., 2001, A\&A,
368, 1021

\bibitem{} Markoff, S., Falcke, H. \& Fender, R., 2001, A\&A, 372L, 25

\bibitem{}McClintock, J.E., et al. 2001, ApJ, 555, 477

\bibitem{} Meier, D.L., 2001, ApJL, 548, 9L

\bibitem{}Menou, K., Hameury, J-M., Lasota, J-P., \& Narayan, R., 2000, MNRAS,
314, 498

\bibitem{}Menou, K., \& McClintock, J.E., 2001, ApJ, 557, 304

\bibitem{}Meyer, F., Liu, B.F. \& Meyer-Hofmeister, E., 2000, A\&A,
361, 175

\bibitem{}Mineshige, S., 1996, PASJ,48,93 

\bibitem{}Mioduszewski, A.J., Rupen, M.P., Hjellming, R.M., Pooley, .G. \&
Waltman, E.B., 2001, ApJ, 553, 766

\bibitem{}Miyamoto, S., Kitamoto, S., Hayashida, K. \& Egoshi, W.,
1995, ApJ, 442, L13

\bibitem{}Muno, M.P., Fox, D.W., Morgan, E.H., \& Bildsten, L., 2000, ApJ, 542, 1016

\bibitem{}Muno, M.P., Remillard, R.A. \& Chakrabarty, D., 2002, ApJL,
568, 35L

\bibitem{}Narayan, R. \& Yi, I., 1994, ApJL, 428, L13

\bibitem{}Narayan, R. \& Yi, I., 1995, ApJ, 452, 710

\bibitem{}Nayakshin S., \& Svensson, R., 2001, ApJL, 551, L67

\bibitem{}Nowak, M.A., 1995, PASP, 107, 1207

\bibitem{}Nowak, M.A., Wilms, J. \& Dove, J.B., 2002, MNRAS, 332, 856

\bibitem{}Orosz, J.A. et al., 2002, ApJ, 568, 845

\bibitem{} Rees, M.J., Phinney, E.S., Begelman, M.C. \& Blandford,
R.D., 1982, Nature, 295, 17

\bibitem{}Reig, P., Mendez, M., van der Klis, M., \& Ford, E.C., 2000, ApJ, 530,916

\bibitem{}Revnitsev, M.G., et al., 2000, MNRAS, 312, 151

\bibitem{}Shakura, N.I., \& Sunyaev, R.A., 1973, A\&A, 24, 337

\bibitem{}Shapiro, S.L., Lightman, A.P. \& Eardley, D.M., 1976, ApJ,
204, 187

\bibitem{}Smith, D.A., Levine, A.M., Remillard, R., Fox, D., \& Schaefer, R., \& RXTE/ASM Team, 2000, IAUC No. 7399

\bibitem{}Smith, D.M., Heindl, W.A. \& Swank, J.H., 2002, ApJ, 569,
362

\bibitem{}Sobczak, G.J., McClintock, J.E., Remillard, R.A., Cui, W., Levine,
A.M., Morgan, E.M., Orosz, J.A., \& Bailyn, C.D., 2000, ApJ, 531, 537

\bibitem{}Srinivasan, G., 1995, in {\it Stellar Remnants}, eds. Meynet \& Schaerer, Springer: Berlin

\bibitem{}Sunyaev, R.A. \& Titarchuk, L., 1980, A\&A, 86, 121

\bibitem{}Svensson, R., "The Non-Linear Phenomena in Accretion Discs
around Black Holes", Reykjavik, June 18-21, 1997, in "Theory of Black
Hole Accretion Discs", eds. M.A. Abramowicz, G. Bjornsson, \&
J.E. Pringle, Cambridge University Press (astro-ph/9902205)

\bibitem{}Tanaka, Y. \& Lewin, W.H.G., 1995, in {\it X-Ray Binaries},
eds. Lewin, van Paradijs, \& van den Heuvel, Cambridge University
Press : Cambridge


\bibitem{}van der Klis, M., 2001, ApJ, 561, 943

\bibitem{}Wilson, C.D. \& Done, C., 2001, MNRAS, 325, 167

\bibitem{}Wilms, J., Nowak, M.A., Dove, J.B., Fender, R.P. \& di
Matteo, T., 1999, ApJ, 522, 460

\bibitem{}Zhang, S.N., Yu, W., \& Zhang, W., 1998, ApJL, 494, L71

\bibitem{}Zhang, W., Jahoda, K., Kelley, R.L., Strohmayer, T.E., Swank J.H., \&
Zhang, S.N., 1998, ApJL, 495, L9

\end{thebibliography}
\end{document}